\documentclass[newstyle,twocolumn,proceedings]{rmaa}
\title{The Galactic Bulge: the Stellar and Planetary Nebulae Populations} 
\author{F. Cuisinier$^1$, J. K\"oppen$^2$, A. Acker$^2$, W.J. Maciel $^3$
\affil{$^1$ Depto. de Astronomia, UFRJ,
       Rio~de~Janeiro, Brazil \\
       $^2$ Observatoire de Strasbourg,
        France \\
       $^3$ Instituto Astron\^omico e Geofisico da USP,
       S\~ao~Paulo, Brazil}}
\fulladdresses{\item email for contact: francois@ufrj.br (F. Cuisinier)}
\shortauthor{Cuisinier et al.}
\shorttitle{The Galactic Bulge: the Stellar and Planetary Nebulae Populations}
\keywords{Planetary Nebulae --- Stars: Red Giants --- Galaxy: (the) Bulge}

\abstract{
We compare abundances patterns in the Bulge for elements observed in stars
and in planetary nebulae. Some $\alpha$ elements, like Mg and Ti, are 
overabundant in respect to Fe,  and others
are not, like He, O, Si, S, Ar, Ca. The first ones favor a quick evolution of 
the Galactic Bulge, and the seconds a much slower one.}

\resumen{
We compare abundances patterns in the Bulge for elements observed in stars
and in planetary nebulae. Some $\alpha$ elements, like Mg and Ti, are
overabundant in respect to Fe,  and others
are not, like He, O, Si, S, Ar, Ca. The first ones favor a quick evolution of
the Galactic Bulge, and the seconds a much slower one.}                    
           
\listofauthors{F. Cuisinier, J. K\"oppen, A. Acker, W.J. Maciel}
\indexauthor{Cuisinier, F.}
\indexauthor{K\"oppen, J.}
\indexauthor{Acker, A.}
\indexauthor{Maciel, W.J.}
\begin{document}

%% This command is necessary to typeset the title, abstract, etc. 
\maketitle

%%
%% And here starts the text....
%%

\section{Introduction}

Color--Magnitude diagrams are nowadays available down to the main sequence in 
the Galactic Bulge with HST, and show without any doubt that stars in the 
 Bulge are old (Gilmore \& Feltzing 2000). However, due to the 
age--metallicity degeneracy,
 it is very hard to make precise age determinations. 
Chemical evolution arguments can here be very helpful.\\

Planetary Nebulae are interesting objects,
because they concentrate the energy of their central stars in the emission
lines of their spectra, and can therefore be observed relatively easily at
this distance. Furthermore the masses of their progenitor stars varying
from 0.8
to  ${\rm 8 M_{\sun}}$,  their ages span from 50 Myr to 25 Gyr,
 covering  more
than 95\% of the possible ages in the Universe, and of course
in the Bulge.\\

The only stars with a similar ages range, where elemental  abundances are  
reasonabily determinable at the distance of the Galactic
Bulge,  are the Red Giants. They  are actually the direct precursors of
the Planetary Nebulae. Some elements have their abundances unmodified
by the stellar evolution in Red Giants as well as in Planetary Nebulae.
These elements keep the fingerprints of the chemical composition of the
ISM when the progenitor star was born, and because of the span of their ages,
they allow to follow its evolution over a very wide time range.\\

One particular point of interest are the relative abundances of elements
produced in type II and in type Ia supernovae. Type II supernovae explode
very rapidely, after some Myr, e.g. quasi instantaneously on
the Bulge evolution timescale, whereas type Ia supernovae explode after
a period of the order of one  Gyr.
The relative abundances of type II and type Ia supernovae should thus
allow to measure the timescale of the Bulge formation.

On the other hand, elements produced during the lifetimes of the progenitor
stars should allow to determine their ages - at least statistically.
In Planetary Nebulae, nitrogen is very easily detectable, and has its
abundance modified in high mass progenitors, that are short lived.
Nitrogen abundances in Planetary Nebulae should thus
help to identify recent star formation.                        

Such a recent star formation is not detected in the stars, but this has to be 
checked in the Planetary Nebulae.
  
\section{Abundances in Stars and in Planetary Nebulae}

We derived  abundances for a sample of 30 PN, that we observed with
high quality spectroscopy (Cuisinier et al. 2000). These abundances
being of really better quality than others available in the
literature, we will only consider these ones here.

Abundances for individual elements in stars are up to now only available
for a sample of 11 red giants, from McWilliam \& Rich (1994).

Unfortunately,  a direct comparison of  abundances is  not possible,
the elements detectable in stars with a good confidence
being different
from those detectable in Planetary Nebulae.\\

We compare therefore the distributions of O, S and Ar in Planetary Nebulae
in the Bulge and in the Disk, these elements representing the pristine
abundances of the ISM (Figure 1, upper panel, for O). We find  the
abundances distributions to be quite similar, like the Fe abundances
in the stars (Mc William \&  Rich 1994).
Furthermore, a comparison of S/O and Ar/O ratios as a function of the O 
abundances shows no particular tendancy (Figure 2), they are fairly constant 
over the whole range of O abundances, at the Disk Planetary Nebulae value.\\

In giant stars, on the other hand, such a comparison shows an 
overabundance of some elements in the Bulge  
in respect to the Disk  tendancies (Mg and Ti), and equivalent abundances
for other elements (Si and Ca) (Mc William \& Rich 1994).\\ 

The N/O ratios comparison in the Bulge and the Disk
(figure 1, lower panel) show that the young progenitor, N-rich Planetary
Nebulae, that are present in the Disk, are lacking in the Bulge. From the
Planetary Nebulae, like from the stars, 
the Bulge does not seem to have formed stars
recently.\\

If the Red Giant and the Planetary Nebulae populations in the
Bulge seem to be quite similar  in the light of our study, the picture
that arises from a comparison of the various  elements originating from
type II and type Ia supernovae that are
detected in Planetary
Nebulae and in Red Giants  remains very puzzling: Mg and Ti, that are enhanced
over Fe,
seem to favor a quick evolution, whereas He, O, Si, S, Ar and Ca show
normal abundances patterns, and favor  a much slower evolution.

\begin{figure}
  \includegraphics[width=\columnwidth]{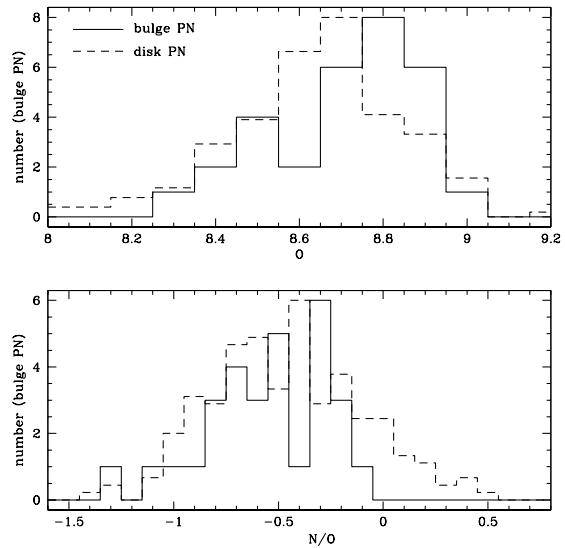}
  \caption{Top panel: oxygen abundances distribution, on a logarithmic scale,
where 12 is the abundance of hydrogen. botom panel: nitrogen over oxygen  
ratios distribution.}
\end{figure}

\begin{figure}
  \includegraphics[width=\columnwidth]{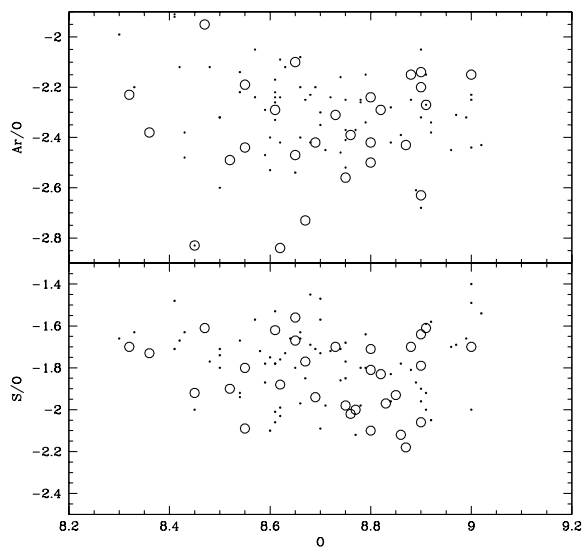}
  \caption{Ar/O and S/O ratios as a function of the oxygen abundance. 
Dots represent Bulge  PN, points Disk type II PN.}
\end{figure}

%\begin{figure*}
%  \includegraphics[width=\textwidth]{rmprocsamp_fig2.ps}
%  \caption{Comparison of morphological predictions of the two-wind
%    interaction model with observations. Contours and greyscales
%    show H$\alpha$ surface brightness (logarithmic scale) for {\em
%      HST} observations and model images. The interval between
%    successive contours is $2^{1/2}$.}
%  \label{fig:comp}
%\end{figure*}

% \begin{figure*}
%  \includegraphics[width=\textwidth]{rmprocsamp_fig3.ps}
%  \caption{Hydrodynamical simulation of the two-wind
%    interaction, calculated in 2D slab symmetry with a grid size of
%    $300\times 300$ cells.  Greyscale shows the gas density while
%    contours show the gas pressure (both logarithmic scale). Arrows
%    show gas velocity. Photoevaporated disk material (white arrows)
% has an isothermal equation of state. Stellar wind material}
%\end{figure*}

% \acknowledgements We are very grateful to Alex Raga, Susana

%% When using the rmaacite package, the \bibitem command should be of
%% the format: 
%%
%%             \bibitem[AUTHOR<YEAR>]{KEY} 
%%
%% so that the \cite{KEY}, etc. commands will work properly. 
%% 
%% If you are doing the citations manually, then you can just use
%% `\bibitem{}' instead. This will give you a warning about
%% `multiply-defined labels' which you can safely ignore.
%% 

\end{document}